\documentclass[12pt]{elsart}
\usepackage{graphicx,ifpdf}
\ifpdf 
\usepackage[pdfauthor={V. Nikolayev},pdftitle={IJHMT01},colorlinks,pdftex]{hyperref} \fi

\begin{document}
\begin{frontmatter}
\title{Growth of a dry spot under a vapor bubble at high heat flux and high pressure}
\author[a1]{V. S. Nikolayev\thanksref{ghi},}
\author[a1]{D. A. Beysens\thanksref{ghi},}
\author[a2]{G.-L. Lagier,}
\and
\author[a3]{J. Hegseth}

\address[a1]{ESEME, Service des Basses Temperatures, CEA Grenoble,
17, rue des Martyrs, 38054, Grenoble Cedex 9, France}
\address[a2]{Laboratoire de Thermohydraulique Avanc\'{e}e, SMTH, CEA Grenoble,
17, rue des Martyrs, 38054, Grenoble Cedex 9, France}
\address[a3]{Department of Physics, University of New Orleans, New Orleans, LA 70148, USA}
\thanks[ghi]{Mailing address: ESEME-CEA, Institut de Chimie de la Mati\`ere Condens\'{e}e de Bordeaux, CNRS,
Avenue du Dr. Schweitzer, 33608 Pessac Cedex, France }
\date\today
\begin{abstract}
We report a 2D modeling of the thermal diffusion-controlled growth of a vapor bubble attached to a heating
surface during saturated boiling. The heat conduction problem is solved in a liquid that surrounds a bubble
with a free boundary and in a semi-infinite solid heater by the Boundary Element Method. At high system
pressure the bubble is assumed to grow slowly, its shape being defined by the surface tension and the vapor
recoil force, a force coming from the liquid evaporating into the bubble. It is shown that at some typical
time the dry spot under the bubble begins to grow rapidly under the action of the vapor recoil. Such a bubble
can eventually spread into a vapor film that can separate the liquid from the heater thus triggering the
boiling crisis (Critical Heat Flux).
\end{abstract}
\end{frontmatter}
\renewcommand{\arraystretch}{1}
\begin{tabular}{|ll|ll|}
\hline
\multicolumn{4}{|c|}{\bf NOMENCLATURE} \\
 \hline $\vec{a}$ &arbitrary vector        &   $x$ &   abscissa \\
 $b$ &   exponent for boundary meshing& $y$ &   ordinate \\
 $C$ &   arbitrary constant & \multicolumn{2}{c|}{\bf Greek letters}\\
 $c_p$   &   specific heat [J/(kg K)]& $\alpha$  &   thermal diffusivity [m$^2$/s]\\
$d_{min}$   &   smallest  element
length [m]&$\beta$& exponent \\
 $\vec{e}_x,\vec{e}_y $  & unit vectors directed along the axes &$\Delta t$   & time step [s]\\
 $F,f$   &non-dimensional time& $\zeta$ &reduced heat flux \\
Fo & Fourier number &$\eta$ & rate of evaporation [kg/(s$\cdot$m$^2$)]\\
$G$ &Green function, BEM coefficient & $\theta$ & liquid contact
angle \\
 $H$ & latent heat [J/kg],  BEM coefficient & $\lambda$ &vapor/liquid pressure difference [N/m$^2$]\\
 Hi & Hickman number & $\xi$& non-dimensional curvilinear coordinate \\
 Ja & Jakob number & $\rho$  &   mass density [kg/m$^3$]\\
  $j$ &volume heat supply [W/m$^3$]&$\sigma$  & surface tension [N/m] \\
 $K$ & curvature [m$^{-1}$] & $\tau$ &dummy\\
 $k$ & thermal conductivity [W/(m K)] & $\psi$  & reduced temperature\\
 $L$& half-length of the bubble contour [m] & $\Omega$& 2D-domain\\
 $M$   & molar weight of water [kg/mol]&$\partial\Omega$ & contour of the 2D-domain \\
 $N$   & total boundary elements number &\multicolumn{2}{c|}{\bf Subscripts} \\
 $\vec{n}$ & internal unit normal vector & $d$  & dry spot \\
 $P_r$ & vapor recoil pressure [N/m$^2$]& $e$  & external to the bubble \\
 $q$ & heat flux [W/m$^2$]& $F,f$ &value at time $F$ or $f$ \\
 $q_{CHF}$ & critical heat flux [MW/m$^2$]& $i$  & vapor-liquid interface \\
 $q_0$   & control value of heat flux [MW/m$^2$] & $i,j$& value at the node $i$ or $j$ \\
 $R$ & bubble radius [m] & $inf$ & at $x\rightarrow\infty$ (also as a superscript) \\
$R_0$ &initial bubble radius [m] & $L$ & liquid \\
$R_g$   & molar gas constant [J/(mol K)]& $max$ & maximum \\
$\vec{r}$ & radius-vector & $S$ & solid (heater)\\
 $T$ & temperature [K]& $sat$  & saturation \\
 $t$ & time [s] & $V$ & vapor \\
 $t_c$ & transition time [s] &$w$&  wetted part of the heater \\
 $t_{dep}$ & bubble residence time [s] &\multicolumn{2}{c|}{\bf Superscripts}  \\
 $u$   & auxiliary angle & $x$ & x-component of the vector \\
 $V$ & 2D-bubble volume [m$^2$]& $y$ & y-component of the vector \\
 $v^n$ &interface velocity [m/s]& $\; \bar{ }$ &reference value \\
 \hline
 \end{tabular}
\renewcommand{\arraystretch}{2}

\section{Introduction}

In nucleate boiling, a very large rate of heat transfer from the heating surface to the bulk is due to both
the phase change (latent heat of vaporization) and the fact that the superheated liquid is carried away from
the heating surface by the departing vapor bubbles. Therefore, the knowledge of the nucleation and growth of
the bubbles on the heating surface is very important for the calculations of the heat transfer rate.  Many
works were focused on the bubble growth kinetics, see e.g. \cite{Tong,Cooper,Stephan,Nydahl,Mei,Dhir}.
However, as it was recently recognized \cite{Buyevich}, the behavior of the fluid in contact with the solid
heater remains poorly studied. We think that this situation is due to the success of the microlayer model
\cite{Cooper} that proved to be self-sufficient for the description of the bubble growth kinetics and the
heat transfer rate. The microlayer model postulates the existence of a thin liquid film between the heater
and the foot of the vapor bubble. This model is based on observations of the bubbles at low system pressures
with respect to the critical pressure for the given fluid. At low pressures, the fast bubble growth creates a
hydrodynamic resistance that makes the bubble almost hemispherical \cite{Carey}. As proved by direct
observations \cite{Van,Tor} through the transparent heating surface, the dry spot (i.e. the spot of the
direct contact between the liquid and the vapor) does exist around the nucleation site while the bubble stays
near the heating surface. The origin of the dry spot can be explained as follows. First, it is necessary that
the vapor-solid adhesion exists to avert the immediate removal of the bubble from the heater by the lift-off
forces. This adhesion only appears when the vapor contacts the solid directly. Second, the strong generation
of vapor at the triple contact line around the nucleation site prevents covering of the nucleation site by
the liquid. As a consequence, the existence of the dry spot under the bubble is necessary during most of the
time of the bubble growth, until the bubble departure from the heater.

Because of the hemispheric bubble shape, the apparent bubble foot is much larger than the dry spot. That is
why the microlayer model works so well at low pressures. For high system pressure, comparable to the critical
pressure (see \cite{Chest,Johnston} for the discussion of the threshold between these two regimes), the
picture is different. The bubble growth is much slower so that the hydrodynamic forces are small with respect
to the surface tension. Consequently, the bubble resembles a sphere much more than a hemisphere \cite{Carey}.
It is very hard to identify the microlayer as a thin film in this case. In this article we will limit
ourselves to this particular case.

One of the most important phenomena in boiling at the large heat fluxes used in industrial heat exchangers is
the boiling crisis called alternately ``burnout'', ``Departure from Nucleate Boiling'', or ``Critical Heat
Flux'' (CHF) \cite{Tong,Carey}. When the heat flux from the heater exceeds a critical value (the CHF), the
vapor bubbles abruptly form a film that thermally insulates the heater surface from the liquid. Consequently,
the temperature of the heater rapidly grows. The dry spot is recognized as playing a key role in the boiling
crisis \cite{Bric}. A new physical approach \cite{EuLet} was recently suggested by some of us to describe
this phenomenon. It is based on the experimental results \cite{Van,Tor} that show that the boiling crisis can
begin with the fast growth of a dry spot under a {\it single} bubble, although several simultaneously
spreading bubbles can coalesce later on. The model associates the onset of the boiling crisis with the
beginning of the spreading of the dry spot below a vapor bubble attached to the heater surface. The purpose
of the present article is to rigorously calculate in 2D the temporal evolution of the dry spot under a single
bubble.

Two main difficulties arise while solving this problem. The first of them is the necessity to solve the full
free-boundary problem for the bubble. Unfortunately, we cannot assume a simple shape for the bubble foot as
has been done in all previous simulations of the vapor bubble growth that we are aware of, see e.g.
\cite{Stephan,Nydahl,Dhir}. The reason is that these models do not rigorously determine the size of the dry
spot. Instead, they use an empirical correlation for the microlayer parameters chosen to satisfy the
experimentally observed growth rate of the bubble. In the present formulation, the dry spot size is
determined in a self-consistent manner from the position of the triple contact line. Such a free-boundary
problem is difficult because of its non-linearity and can be solved by only a few numerical methods, e.g.
Boundary Element Method (BEM) \cite{Breb} or Front Tracking Method. The latter was recently used \cite{Tryg}
to simulate the film boiling in 2D.

The second difficulty of the dry spot problem is associated with the calculation of the heat transfer in the
most important region --- the vicinity of the triple contact line (i.e. the microlayer). It is analyzed
analytically in the subsection \ref{ssw} for two fixed values of the contact angle. Although these results
are not used in the numerical simulation, we need them to check the accuracy of the heat transfer calculation
in this important region.

 Our calculations are valid in the simplified situation where the hydrodynamic effects in liquid are
neglected. This simplification is justified by the slow growth assumption valid for high system pressures.
This approximation is common \cite{Mei} for the thermal diffusion-controlled bubble growth. The only dynamic
condition that cannot be neglected \cite{EuLet} is the dynamic balance of mechanical momentum at the bubble
interface that results in the vapor recoil pressure. This approach allows us to apply the quasi-static
approximation for the bubble shape determination in sec.~\ref{S3} and neglect the convection terms in the
heat transfer problem that is discussed in sec.~\ref{sec5}. The numerical algorithm is described in section
\ref{sec6}. The results of the simulation and the conclusions are presented in sections \ref{sec7} and
\ref{sec8}.

\section{Bubble shape determination}\label{S3}

When the heat flux $q_S$ from the heater is small, the vapor bubble grows with its triple vapor-liquid-solid
contact line pinned by the surface defect on which the vapor nucleation has started. The size of the dry spot
is thus very small with respect to the bubble size. According to the model \cite{EuLet} that should be valid
for any system pressure, at some value of $q_S$ the contact line depins and spreads under the influence of
the vapor recoil pressure $P_r$,
\begin{equation}
P_r=\eta^2(\rho_V^{-1}-\rho_L^{-1}), \label{Pr}
\end{equation}
where $\eta$ is the mass of the evaporated liquid per unit time per unit area of the vapor-liquid interface.
$P_r$ may vary along the interface and is directed normally to this interface towards the liquid
\cite{Plesset}. By neglecting heat conduction in the vapor with respect to the latent heat effect, $\eta$ can
be related to the local heat flux across the interface $q_L$ by the equation
\begin{equation} q_L=H\eta, \label{eta}
\end{equation} where $H$ is the latent heat of vaporization.
Because of the strong temperature gradient in the vicinity of the heating surface, $q_L$ increases sharply
near the contact line, and consequently $\eta$ and $P_r$ also increase. In other words, the pressure
increases near the contact line and causes it to recede. Therefore, the dry spot under the bubble should grow
with time.

The bubble shape is determined using a quasi-static approximation. We neglect all but two forces that define
the bubble shape: the surface tension and the vapor recoil pressure $P_r$ defined by Eq.~\ref{Pr}. The bubble
shape is then defined by the pressure balance (see \cite{EuLet})
\begin{equation} K\sigma=\lambda+P_r,\label{surf}
\end{equation}
where $K$ is the local curvature of the bubble, $\sigma$ is the vapor-liquid interface tension and $\lambda$
is a constant difference of pressures between the vapor and the liquid. $\lambda$ should be determined using
the known volume $V$ of the 2D bubble. At any time, the volume $V$ of the 2D bubble (see Appendix \ref{A1})
can be written as:
\begin{equation}
V={1\over 2}\int\limits_{(\partial\Omega_i)} (xn_e^x+yn_e^y)
\;{\rm d}\,\partial\Omega,\label{V}
\end{equation}
where $\partial\Omega_i$ is the vapor-liquid interface, and the external unit normal vector
$\vec{n}_e=(n_e^x,n_e^y)$ to the bubble is defined in a Cartesian $(x,y)$ coordinate system. An axisymmetric
bubble shape is assumed in the following, see Fig.~\ref{bubble}.

It is convenient to describe the bubble shape in parametric form while choosing a non-dimensional length
$\xi$ measured along the bubble contour as an independent variable. Then the coordinates $(x,y)$ for a given
point on the bubble interface are functions of $\xi$ that varies along the bubble half-contour (its right
half in Fig.~\ref{bubble}) from 0 to 1, $\xi=0$ and $\xi=1$ corresponding to the topmost point of the bubble
and to the contact point respectively. Eq.~\ref{surf} for the 2D case is equivalent to the following
parametric system of ordinary differential equations:
\begin{eqnarray}
{\rm d}x/{\rm d}\xi  &=&L\cos u,  \label{vol1} \\ {\rm d}y/{\rm
d}\xi  &=&-L\sin u,  \label{vol2} \\ {\rm d}u/{\rm d}\xi
&=&L(\lambda +P_{r}(\xi ))/\sigma   \label{vol3}
\end{eqnarray}
were $u=u(\xi)$ is the angle between the tangent to $\partial\Omega_i$ at the point $\xi$ and the vector
directed opposite to the $x$-axis; $L$ is the half-length of $\partial\Omega_i$.

The boundary conditions for Eqs. \ref{vol1} -- \ref{vol3} are then given by
\begin{equation}
x(0)=0, \quad u(0)=0, \quad y(1)=0.\label{bound}
\end{equation}
A 4th condition $u(1)=\pi-\theta$ that fixes the liquid contact angle $\theta$ is necessary to determine the
unknown $L$ using (\ref{vol3}): \begin{equation} L=(\pi-\theta)\sigma\left[\int\limits_0^1 P_r(\xi) {\rm d}\xi
+\lambda\right]^{-1}.\label{L}
\end{equation}
In the following, we consider the usual case of complete wetting of the heating surface by the liquid,
$\theta=0$. The solution of the problem (\ref{Pr} -- \ref{L}) allows the bubble shape to be determined
providing the heat flux through the vapor-liquid interface is known.

\section{Heat transfer problem}\label{sec5}

The calculation of the heat transfer around a growing vapor bubble is a free boundary problem. To our
knowledge, only two other groups have solved the full free-boundary boiling problem \cite{Tryg,Mazouzi}. In
both works the singular effects, which appear in the region adjacent to the triple contact line (microlayer),
are not discussed. However, we know that almost all the heat flux supplied to the vapor bubble goes through
this particular region, a region on which we concentrate in this work. As a first step, we neglect the heat
transfer due to the liquid motion. Thermal conduction and the latent heat effects are taken into account to
describe the time evolution of the 2D vapor bubble. The vapor is assumed to be non-conducting. The case of
saturated boiling is considered. This means that the temperature in the liquid far from the heater is equal
to $T_{sat}$, the saturation temperature for the given system pressure. Thus only evaporation is allowed on
the bubble interface. Since we consider a slow process with no liquid motion, the pressures are assumed to be
uniform both in the vapor and in the liquid. However, they are different according to (\ref{surf}) where
$\lambda$ is the difference between the pressures inside and outside the vapor bubble. The saturation
temperature $T'_{sat}$ for the vapor inside the bubble depends on the vapor pressure according to the
Clausius-Clapeyron equation (see, e.g. \cite{Nydahl}) $$T'_{sat}=T_{sat}[1+{\lambda\over
H}(\rho_V^{-1}-\rho_L^{-1})].$$ Since $\sigma/\lambda$ is the bubble radius at the top of the bubble (where
$P_r$ is negligible), it is easy to estimate that the correction to $T_{sat}$ is less than $10^{-3}$\% even
for the smallest bubble size considered. Therefore, in the following, the bubble surface is supposed to be at
constant temperature $T_{sat}$.

\subsection{Model for the vicinity of the contact line}\label{ssw}

First of all, we need to understand how $q_L$ behaves in the vicinity of the contact line where the contour
of the bubble $\partial\Omega_i$ can be approximated  by a straight line that forms an angle $\theta$ with
the $Ox$ heater line, see Fig.~\ref{wedge}. Then $q_L$ can be obtained from the solution of a simple
two-dimensional problem of unsteady heat conduction in this wedge of liquid, the point $O(x=0, y=0)$
corresponding to the contact line. In our previous article \cite{EuLet} the model problem was solved to show
that when $\theta=\pi/2$ and a constant heat flux from the heater is imposed, the heat flux through the
liquid-vapor interface $q_L$ diverges weakly (logarithmically) near the contact line. In the present work we
treat two cases $\theta=\pi/4$ and $\theta=\pi/8$ which also allow an analytical treatment.

The heat conduction equation for the temperature $T_L(x,y,t)$ in the liquid
\begin{equation}
{\partial T_L\over\partial t}=\alpha_L \nabla^2 T_L\label{eqnL}
\end{equation}
has the initial and boundary conditions
\begin{eqnarray}
\left.T_L\right|_{t=0}=T_{sat}, \label{init}\\
\left.T_L\right|_{\partial\Omega_i} =T_{sat},\label{Tsat}\\
-k_L\left.{\partial T_L\over\partial
y}\right|_{y=0}=q_0,\label{qconst}
\end{eqnarray}
where $k_L$ and $\alpha_L$ are the thermal conductivity and the thermal diffusivity of the liquid, and the
heat flux $q_S$ from the heating surface is assumed to be constant ($=q_0$) for the case of the thin heating
wall. The solution of this 2D problem for the angles $\theta=\pi/2^m$, where $m$ is integer, can be obtained
using the method of images \cite{Carslaw} for the Green function. For $\theta=\pi/4$ it reads
\begin{equation}
T_L=T_{sat}+T_{inf}(y,t)-T_{inf}(x,t), \label{T4}
\end{equation}
where the function $T_{inf}(y,t)$ is a solution for this problem at $x\rightarrow\infty$
\begin{equation}
T_{inf}(y,t)= {q_0\over k_L}\left[\sqrt{4\alpha_L
t\over\pi}\exp\left(-{y^2\over 4\alpha_L t}\right)-
y\,\mbox{erfc}\left({y\over 2\sqrt{\alpha_Lt}}\right)\right],
\label{Tinf}
\end{equation}
erfc$(z)$ being the complementary error function \cite{Abr}. Then the heat flux $q_L$
\begin{equation} q_L=-k_L(\vec{n}\cdot\nabla) \left.T_L\right|_{\partial\Omega_i} \label{qL}
\end{equation}
can be calculated as a function of $x$ using the expression for
the unit normal vector $\vec{n}=(-\sin\theta,\, \cos\theta)$:
\begin{equation} q_L= q_0\sqrt{2}\,\mbox{erfc}\left({x\over 2\sqrt{\alpha_Lt}}\right). \label{qL4}
\end{equation}
It is easy to see that, unlike the problem \cite{EuLet} for $\theta=\pi/2$, the heat flux remains finite at
the contact line ($x=0$). This is also true for the case $\theta=\pi/8$, for which
\begin{equation}
T_L=T_{sat}+T_{inf}(y,t)+T_{inf}(x,t)-T_{inf}\left({x+y\over\sqrt{2}},t\right)-
T_{inf}\left({x-y\over\sqrt{2}},t\right), \label{T8}
\end{equation}
and
\begin{equation} q_L=q_0\sqrt{\sqrt{2}+2}\left[\mbox{erfc}
\left({(\sqrt{2}-1)x\over 2\sqrt{\alpha_Lt}}\right) -(\sqrt{2}-1)\,
\mbox{erfc}\left({x\over 2\sqrt{\alpha_Lt}}\right)\right]. \label{qL8}
\end{equation}

Although these solutions present important benchmarks for the numerical calculations of the heat transfer
near the contact line, they cannot be used in the simulation itself for two reasons. The first is the
impossibility to employ the uniform heat flux boundary condition (\ref{qconst}) because in reality the heat
flux vary strongly between the dry and wetted parts of the solid surface. In particular, the heat flux
through the dry spot under a bubble is very small (we assume it to be zero in the following). The second
reason is the impossibility to approximate the bubble contour by a straight line for the case $\theta=0$,
most important for the industrial applications.

\subsection{Mathematical formulation}

We consider the growth of a vapor bubble on the semi-infinite ($y<0$) solid heater $\Omega_S$ in the
semi-infinite ($y>0$) liquid $\Omega_L$, see Fig.~\ref{bubble}. We assume that no superheat is needed for the
bubble nucleation so that the circular bubble of the radius $R_0$ and the volume $V_0$ has already nucleated
at the heater surface at $t=0$. The validity of this assumption is discussed in sec.~\ref{sec7}. The known
heat supply $j(t)$ is generated homogeneously inside the heater with the heat conductivity $k_S$ and the heat
diffusivity $\alpha_S$ so that heat conduction equation for the domain $\Omega_S$
\begin{equation}
{\partial T_S\over\partial t}=\alpha_S \nabla^2 T_S+{\alpha_S\over
k_S}j(t), \quad y<0 \label{eqnS}
\end{equation}
should be solved with the boundary and the initial conditions
\begin{eqnarray}
q_S=-k_S\left.{\partial T_S\over\partial y}\right|_{y=0}=\left\{
\begin{array}{lll}
\displaystyle -k_L\left.{\partial T_L\over\partial y}\right.&
\mbox{at} &\partial\Omega_w\\ 0& \mbox{at} &\partial\Omega_d
\end{array}\right.,
 \label{qS0}\\
\left.T_S\right|_{\partial\Omega_w}=\left.T_L\right|_{y=0},
\label{TS0}\\ \left.T_S\right|_{t=0}=T_{sat}, \label{initS}
\end{eqnarray}
where $\partial\Omega_d$ is the vapor-solid interface (i.e. the dry spot) and $\partial\Omega_w$ is the
liquid-solid interface(wetted surface), see Fig.~\ref{bubble}. The problem for the domain $\Omega_L$ is
completed by Eqs. \ref{eqnL} -- \ref{Tsat}.

The bubble volume $V$ increases due to evaporation \cite{Tong}:
\begin{equation}
H\rho_V{{\rm d}V\over{\rm d}t}=\int\limits_{(\partial\Omega_i)}
q_L \;{\rm d}\,\partial\Omega, \label{eqnV}
\end{equation}
where $q_L$ is calculated using (\ref{qL}) in which $\vec{n}$ is the {\em inner} normal vector to
$\partial\Omega_i$, $\vec{n}=-\vec{n}^e$.

The formulated mathematical problem can be solved by the Boundary Element Method (BEM) generalized for moving
boundary problems \cite{Wrobel}. Before its direct application, we consider the integration contours for the
domains $\Omega_L$ and $\Omega_S$. Obviously, they should contain $\partial\Omega_i$, $\partial\Omega_d$ and
$\partial\Omega_w$ and a circle that closes the contours at infinity.  The non-zero values of the temperature
and flux at infinity complicates the solution. Therefore, we calculate these values $T_S^{inf}$ and
$T_L^{inf}$ at $x\rightarrow\infty$ and then subtract them from  $T_S$ and $T_L$. The resulting modified
variables are zero at infinity. This transformation allows the integration contour to be reduced to
$\partial\Omega_i\cup\partial\Omega_d\cup\partial\Omega_w$.

\subsection{Solution at infinity}

The solution at infinity satisfies the same problem as $T_S$ and $T_L$ but with the eliminated dependence on
$x$ and $\partial\Omega_d=\emptyset$. The separate solutions for $y\ge 0$ and $y\le 0$ can be easily found
using the known Green function for the semi-infinite space \cite{Carslaw}:
\begin{eqnarray}
T_L^{inf}=T_{sat}+{\sqrt{\alpha_L}\over k_L\sqrt{\pi}}\int\limits_0^t{q_0(t-\tau)\over
\sqrt{\tau}}\exp\left(-{y^2\over 4\alpha_L t}\right)\,{\rm d}\tau,\quad y\ge 0\label{TLinf}\\
T_S^{inf}= T_{sat}+{\alpha_S\over k_S}\int\limits_0^tj(\tau)\,{\rm d}\tau-{\sqrt{\alpha_S}\over
k_S\sqrt{\pi}}\int\limits_0^t{q_0(t-\tau)\over \sqrt{\tau}}\exp\left(-{y^2\over 4\alpha_S t}\right)\,{\rm
d}\tau,\quad y\le 0\label{TSinf}
\end{eqnarray}
The unknown heat flux from the heater, $q_0(t)$, can be found for arbitrary $j(t)$ out of the integral
equation that results from equality of (\ref{TLinf}) and (\ref{TSinf}) at $y=0$. It is worth mentioning that
if $j\propto t^{-1/2}$, a constant $q_0$ satisfies this integral equation. This means that in the bubble
growth problem with this choice of $j(t)$ the heat flux from the heater would remain constant at least far
from the growing bubble. This choice will allow us to avoid the influence of the varying heat flux on the
bubble growth and thus will be used in the following. The solution in analytical form is:
\begin{eqnarray}
j(t)=C/\sqrt{t},\nonumber\\ q_0=C\sqrt{\pi}\alpha_S
k_L/(k_S\sqrt{\alpha_L}+k_L\sqrt{\alpha_S}), \nonumber\\
T_S^{inf}= T_{sat}+{2\alpha_S\over k_S}C\sqrt{t}-{q_0\over k_S}\left[\sqrt{4\alpha_S
t\over\pi}\exp\left(-{y^2\over 4\alpha_S t}\right)+ y\,\mbox{erfc}\left(-{y\over
2\sqrt{\alpha_St}}\right)\right], \quad y\le 0,\label{TSinf0}\\
T_L^{inf}=T_{sat}+T_{inf}(y,t),\quad y\ge 0,\label{TLinf0}
\end{eqnarray}
where $C$ and $q_0$ are constant and the function $T_{inf}(y,t)$ is defined in (\ref{Tinf}). It is easy to
see that besides the advantage of the zero values at infinity for the modified variables
$T_{L,S}-T_{L,S}^{inf}$, the equation for both of them has the form (\ref{eqnL}) with no source term.

\subsection{Non-dimensional formulation}

By introducing the characteristic scales for time ($\Delta t$, the time step), length ($R_0$, the initial
bubble radius), heat flux ($\bar{q}$), and thermal conductivity ($\bar{k}$), all other variables can be made
non-dimensional. In particular, the characteristic temperature scale in the system is $\bar{q}R_0/\bar{k}$.
The following four non-dimensional groups define completely the behavior of the system
\begin{eqnarray*}
{\rm Fo}_{L,S}=\alpha_{L,S}\Delta t/R_0^2 \mbox{ --- Fourier
numbers},\\ {\rm
Ja}={\rho_Lc_{p\,L}\over\rho_VH}{\bar{q}R_0\over\bar{k}} \mbox{
--- Jakob number\cite{Tong}},\\ {\rm Hi}={R_0\,\bar{q}^2\over\sigma
H^2}(\rho_V^{-1}-\rho_L^{-1}) \mbox{
--- Hickman number},
\end{eqnarray*}
providing that non-dimensionalized values of $q_0$ and $k_{L,S}$ are fixed. The following is the complete
non-dimensional heat transfer problem formulated in terms of $\psi_{L,S}=(T_{L,S}-T_{L,S}^{inf})/(\bar{q}R_0/
\bar{k})$:
\begin{eqnarray}
{\partial \psi_{L,S}\over\partial t}={\rm Fo}_{L,S}\nabla^2
\psi_{L,S}\label{eqnP}\\ \left.\psi_{L,S}\right|_{t=0}=0,
\label{initP}\\ \left.\psi_L\right|_{\partial\Omega_i}
=-T_{inf}(y,t), \label{bound1} \\
\left.\zeta_L\right|_{\partial\Omega_w}=\left.\zeta_S\right|_{\partial\Omega_w},
\quad
\zeta_{L,S}=k_{L,S}{\partial\psi_{L,S}\over\partial\vec{n}},\label{bound2}
\\ \left.\zeta_L\right|_{\partial\Omega_d}=-q_0\label{bound3} \\
{{\rm d}V\over{\rm d}t}={\rm Fo}_L\cdot{\rm
Ja}\int\limits_{(\partial\Omega_i)} (\zeta_{inf}-\zeta_L) \;{\rm
d}\,\partial\Omega,\quad\zeta_{inf}=q_0n^y\,\mbox{erfc}\left({y\over
2\sqrt{{\rm Fo}_L t}}\right)\quad\label{Vn}
\end{eqnarray}
where $n^y$ is the ordinate of the vector $\vec{n}$ and all quantities are non-dimensionalized. The whole
problem is completed using the non-dimensionalized set of equations for the bubble shape (\ref{vol1} --
\ref{L}) where the non-dimensional expression for the vapor recoil pressure is used:
\begin{equation}
P_r={\rm Hi}(\zeta_{inf}-\zeta_L)^2\label{Prn}.
\end{equation}

\subsection{Boundary Element techniques applied to bubble growth}

As it is shown in \cite{Wrobel}, the heat conduction problem
(\ref{eqnP} -- \ref{initP}) is equivalent to the set of two
integral equations, written for each of the domains $\Omega_L$ and
$\Omega_S$:
\begin{eqnarray}
\int\limits_{0}^{t_F}{\rm
d}t\int\limits_{(\partial\Omega_{L,S})}\Biggl[G^{L,S}(\vec{r'},t_F;\vec{r},t)
\left({\rm Fo}_{L,S}{\zeta(\vec{r},t)\over
k_{L,S}}+\psi(\vec{r},t) v^n(\vec{r},t) \right)- \nonumber\\
\left.{\rm Fo}_{L,S}\, \psi(\vec{r},t){\partial_r
G^{L,S}(\vec{r'},t_F;\vec{r},t)\over\partial\vec{n}} \right]{\rm
d}_r\partial\Omega={1\over 2}\psi(\vec{r'},t_F),\label{Ieq}
\end{eqnarray}
where $\vec{r'}$ is the evaluation point and $t_F$ is the evaluation time. The integration is performed over
the closed contours $\partial\Omega_L$ and $\partial\Omega_S$ that surround the domains $\Omega_L$ and
$\Omega_S$, vector $\vec{n}$ being external to them. $v^n$ is the projection of the local velocity of the
possibly moving integration contour on the vector $\vec{n}$. Since the points $\vec{r'}$ and $\vec{r}$ belong
to these contours, the BEM formulation does not require the values of $\psi$ and $\zeta$ to be calculated in
the internal points of the domains, which is a great advantage of this method. The functions $G^{L,S}$ are
the Green functions for the equations \cite{Breb}, adjoint to (\ref{eqnP}):
\begin{equation}
G^{L,S}(\vec{r'},t_F;\vec{r},t)={1\over 4\pi{\rm Fo}_{L,S}(t_F-t)}\exp\left[-{|\vec{r'}-\vec{r}|^2\over
4\,{\rm Fo}_{L,S}(t_F-t)}\right].\end{equation}

The indices $L$ and $S$ will be dropped for the sake of clarity
until the end of this section.

The constant element BEM \cite{Breb} was used, i. e. $\zeta$ and
$\psi$ were assumed to be constant during any time step and on any
element, their values on the element being associated with the
values on the node at the center of the element. The time steps
are equal, i.e. $t_f=f$. Therefore, the values of $\zeta$ and
$\psi$ on the element $j$ at time $f$ can be denoted by
$\zeta_{fj}$ and $\psi_{fj}$. Each of the integral equations
(\ref{Ieq}) reduces to the system of linear equations
\begin{equation}
\sum\limits_{f=1}^F\sum\limits_{j=1}^{2N_f}[(\zeta_{fj}/k+ \psi_{fj}v^n_{fj}/{\rm
Fo})G_{ij}^{Ff}-\psi_{fj}H_{ij}^{Ff}]=\psi_{Fi}/2,\label{lin1}
\end{equation}
where $N_f$ is the number of elements on one half of the integration contour at time step $f$, $F_{max}$ is
the maximum number of time steps for the problem; $i=1\ldots 2N_F$ and $F=1\ldots F_{max}$. It is important
that the algorithm for the calculation of the coefficients $H_{ij}$ and $G_{ij}$ \cite{Breb} be fast. We used
the analytical expressions calculated \cite{Lagier} for the case $i=j$. For all other cases the coefficient
$H_{ij}$ can be expressed analytically \cite{Lagier} through $G_{ij}$. $G_{ij}$ was calculated numerically.
The system (\ref{lin1}) can be simplified due to axial symmetry of the problem ($\psi_{fj}=\psi_{f(2N_f-j)}$,
etc.):
\begin{equation}
\sum\limits_{f=1}^F\sum\limits_{j=1}^{N_f}[(\zeta_{fj}/k+ \psi_{fj}v^n_{fj}/{\rm
Fo})\tilde{G}_{ij}^{Ff}-\psi_{fj}\tilde{H}_{ij}^{Ff}]=\psi_{Fi}/2,
\end{equation}
where $i=1\ldots N_F$, $F=1\ldots F_{max}$, $\tilde{G}_{ij}^{Ff}=G_{ij}^{Ff}+G_{i(2N-j)}^{Ff}$ and
$\tilde{H}_{ij}^{Ff}=H_{ij}^{Ff}+H_{i(2N-j)}^{Ff}$. This equation can be rewritten in the form that reveals
explicitly the unknown variables on each time step $F$:
\begin{eqnarray}
\sum\limits_{j=1}^{N_F}[(\zeta_{Fj}/k+ \psi_{Fj}v^n_{Fj}/{\rm
Fo})\tilde{G}_{ij}^{FF}-\psi_{Fj}(\tilde{H}_{ij}^{FF}+1/2)]=\nonumber\\
-\sum\limits_{f=1}^{F-1}\sum\limits_{j=1}^{N_f}[(\zeta_{fj}/k+
\psi_{fj}v^n_{fj}/{\rm
Fo})\tilde{G}_{ij}^{Ff}-\psi_{fj}\tilde{H}_{ij}^{Ff}].\label{lin2}
\end{eqnarray}
Unfortunately, no effective time marching scheme \cite{Breb} can be applied because of the free boundaries.
Since the terms in the sum over $f$ decrease with the decrease of $f$, this sum can be truncated as suggested
in \cite{Demirel}. However, in our case, the magnitude of these terms can be controlled directly because the
coefficients $H$ and $G$ must be recalculated for each $f$. It should be noted that, because of moving
boundaries, the positions of the $i$-th point at times $f$ and $F$ can be different. Therefore, it is very
important that $\tilde{G}_{ij}^{Ff}$ be calculated using the coordinates of the $i$-th point at time moment
$F$ and those of $j$-th point at time moment $f$.

\subsection{Validation of the algorithm for BEM }

\begin{figure}[ht]\centering\includegraphics[width=8cm]{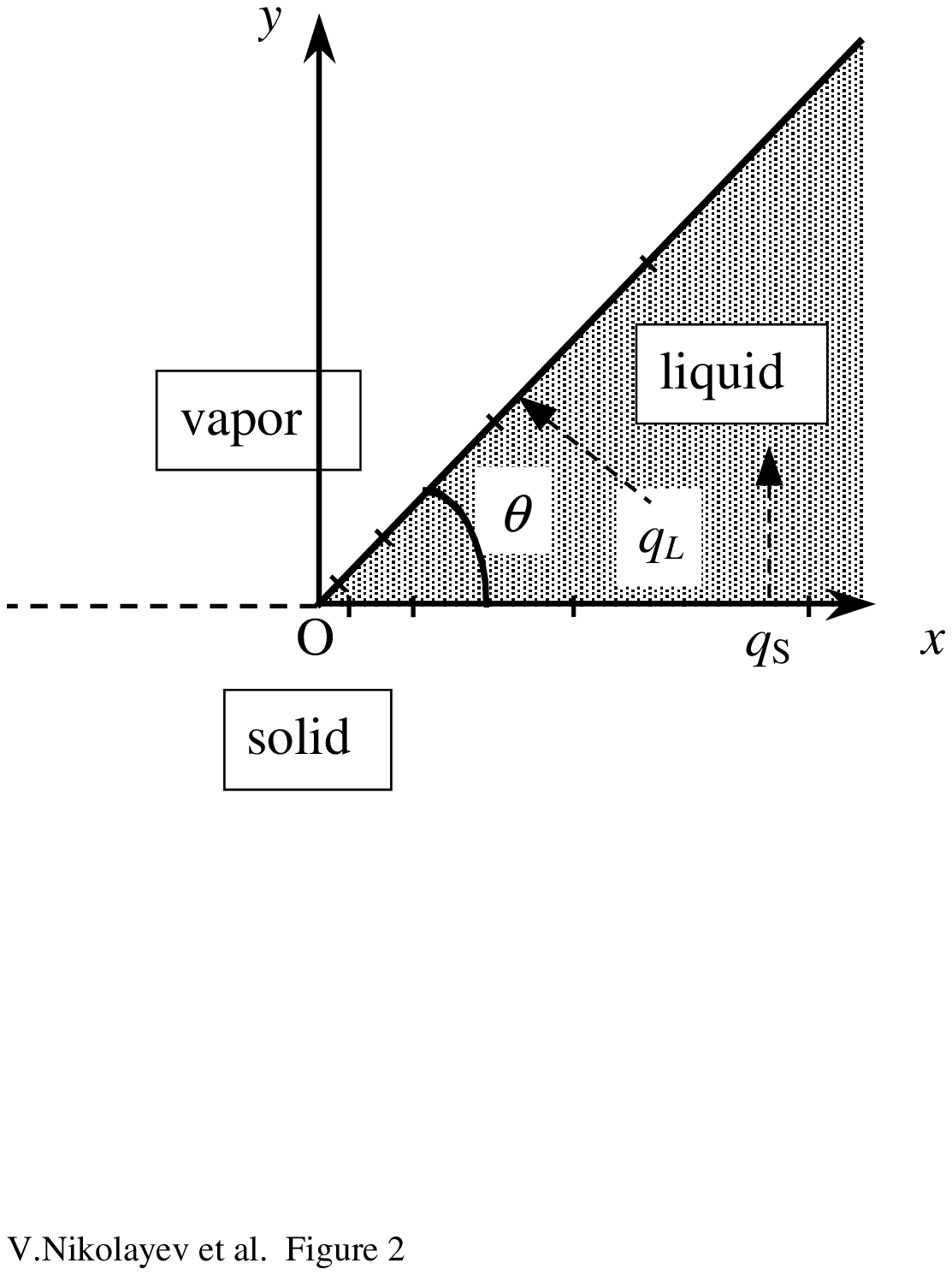}
\caption{Geometry for the analytical calculation of the heat conduction in the wedge. The BEM discretization of
the wedge is also illustrated.}\label{wedge}
\end{figure}

The BEM algorithm was tested for the wedge problem solved analytically in subsection \ref{ssw}. The adaptive
discretization of the integration contour is organized as follows. Since $\zeta$ decreases to zero far from
the bubble, the two ending points (most distant from the contact point $(0,0)$) can be found for the given
$t$ from the condition that $\zeta(x,y,t)$ be sufficiently small. In practice, $x_{max}\sim 10\sqrt{{\rm
Fo}\,t}$. The element lengths grow exponentially ($d_{min}, d_{min}e^b, d_{min}e^{2b},\ldots$) from the
contact point into each of the sides of the wedge (see Fig.~\ref{wedge}), where $b$ is fixed at 0.2. Being an
input parameter, $d_{min}$ is adjusted slightly on each time step to provide the exponential growth law for
the elements on the interval with the fixed boundaries $(0, x_{max})$. Since $x_{max}$ increases, the total
number of the elements also increases during the evolution of the bubble. Remeshing on each time step was
performed to comply with the free boundary nature of the main problem where the remeshing is mandatory.

\begin{figure}[ht]\centering\includegraphics[width=8cm]{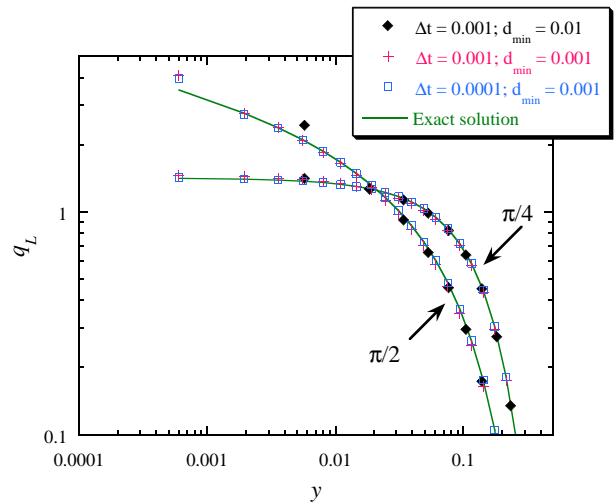}
\caption{The $q_L(y)$ curves calculated for the $\pi/2$ and $\pi/4$ wedges and for the values of the parameters
$q_0=1$, $\alpha_L=1$ and $t=0.01$. The results of the numerical solution by BEM (to be compared with the exact
analytical solution) are presented for the different time and space discretization parameters.} \label{test}
\end{figure}

The results for $\theta=\pi/4$ and $\pi/2$ are shown in Fig.~\ref{test} to be compared with the solid curves
calculated using (\ref{qL4}) and its analog for $\theta=\pi/2$ (Eq.~8 from \cite{EuLet}). It is easy to see
that the method produces excellent results even for coarse discretization, except for the element closest to
the contact point. The algorithm overestimates the value of $q_L$ at this element. The error is larger for the
$\pi/2$ wedge, for which $q_L\rightarrow\infty$ at the contact point. Fig.~\ref{test} demonstrates that the
increase of the numerical error with the increase of the time and space steps is very weak.

\section{Numerical implementation}\label{sec6}

Since we chose $\psi$ and $\zeta$ to be zero at infinity, (\ref{Ieq}) is satisfied trivially on the
semicircles of the infinite radius that close the contours $\partial\Omega_L$ and $\partial\Omega_S$. Thus
these circles can be excluded. Then $\partial\Omega_L=\partial\Omega_i\cup\partial\Omega_w$ and
$\partial\Omega_S=\partial\Omega_d\cup\partial\Omega_w$. The direction of the unit normal vector $\vec{n}$ is
chosen to be external to $\Omega_L$  in the following, see Fig. \ref{bubble}. Then it is internal to
$\Omega_S$, which requires the sign of the integral over $\partial\Omega_S$ to be changed. Making use of the
boundary conditions (\ref{bound1} -- \ref{bound3}), the system of Eqs. \ref{Ieq} reduces to
\begin{eqnarray}
\int\limits_{0}^{t_F}{\rm
d}t\Biggl\{\;\int\limits_{(\partial\Omega_i)}\left[ G^L\left({\rm
Fo}_L{\zeta_L\over k_L} -T_{inf} v^n\right)+{\rm Fo}_L
T_{inf}{\partial G^L\over\partial\vec{n}} \right]{\rm
d}\partial\Omega+\nonumber\\ {\rm
Fo}_L\int\limits_{(\partial\Omega_w)}\left(G^L{\zeta_S\over k_L}
-\psi_S {\partial G^L\over\partial\vec{n}} \right){\rm
d}\partial\Omega \Biggr\}={1\over 2}\left\{\begin{array}{ll}
\psi_S,&\vec{r}_F\in\partial\Omega_w,\\
-T_{inf},&\vec{r}_F\in\partial\Omega_i,
\end{array}\right.
 ,\label{Ieq1}\\
{\rm Fo}_S \int\limits_{0}^{t_F}{\rm d}t\Biggl[\;\int\limits_{(\partial\Omega_w)}\left(-G^S {\zeta_S\over
k_S}+\psi_S{\partial G^S\over\partial\vec{n}} \right){\rm d}\partial\Omega+\nonumber\\
\int\limits_{(\partial\Omega_d)}\left(G^S{q_0\over k_S} +\psi_S {\partial G^S\over\partial\vec{n}}
\right){\rm d}\partial\Omega \Biggr]= {1\over 2}\psi_S, \quad\vec{r}_F\in\partial\Omega_d\cup\partial\Omega_w,
 \label{Ieq2}
\end{eqnarray}
where the arguments of all functions are supposed to be exactly as in (\ref{Ieq}). These equations should be
solved using the BEM described in the previous section for unknown functions $\zeta_L(\vec{r},t)$ for
$\vec{r}\in\partial\Omega_i$, $\psi_S(\vec{r},t)$ for $\vec{r}\in
\partial\Omega_d\cup\partial\Omega_w$, and $\zeta_S(\vec{r},t)$
for $\vec{r}\in\partial\Omega_w$.

The discretization of the integration subcontours $\partial\Omega_w$, $\partial\Omega_d$ and
$\partial\Omega_i$ follows the same exponential scheme (see Fig. \ref{bubble}) that was used for the
discretization of the wedge sides in the test example above. The only difference is the axial symmetry of the
mesh that corresponds to the symmetry of the bubble. Since the free boundary introduces a nonlinearity into
the problem, the following iteration algorithm is needed to determine the bubble shape on each time step
\cite{Wrobel}:
\begin{enumerate}
  \item Shape of the bubble is guessed to be the same
  as on the previous time step;
  \item The variations of $v^n$ and $P_r$ along the bubble
  interface are guessed to be the same as on the previous time step;
  \item Discretization of the contours $\partial\Omega_w$,
  $\partial\Omega_d$ and $\partial\Omega_i$ is performed;
  \item Temperatures and fluxes on the contours $\partial\Omega_w$,
  $\partial\Omega_d$ and $\partial\Omega_i$ are found using the described above BEM
techniques;
  \item Volume $V$ and vapor recoil $P_r$ are calculated using
  (\ref{Vn}) and (\ref{Prn});
  \item Bubble shape is determined (see Sec.~\ref{S3}) for the
  calculated values of $V$ and $P_r$;
  \item If the calculated shape differs too much from that on the
  previous iteration,  the velocity of interface $v^n$ is calculated, and
   steps 3 -- 7 are repeated until the required accuracy is attained.
\end{enumerate}
As a rule, three iterations give the 0.1\% accuracy which is sufficient for our purposes.

The normal velocity of interface $v^n_{Fi}$ at the time $F$ and at node $i$ is calculated using the expression
\begin{equation}\label{vn0}
  v^n_{Fi}=(x_{Fi}-x_{(F-1)j})n^x_{(F-1)j}+(y_{Fi}-y_{(F-1)j})n^y_{(F-1)j},
\end{equation}
where $x_{Fi}$ is the coordinate of the node $i$ at time $F$, and $j$ is the number of the node (at time
$F-1$) geometrically closest to $(x_{Fi},y_{Fi})$.

The system of Eqs. \ref{vol1} -- \ref{vol3} is solved by direct integration. The integration of the
right-hand side of (\ref{vol3}) is performed using the simple mid-point rule, because the values of $P_r$ are
calculated at the mid-points (nodes) only. The subsequent integration of the right-hand sides of
Eqs.~\ref{vol1} -- \ref{vol2} is performed using the Simpson rule (to gain accuracy) for the non-equal
intervals. The trapezoidal rule turns out to be accurate enough for the calculation of volume in (\ref{V}).
For the simulation we used the parameters for water at 10 MPa pressure on the heater made of stainless steel
(Table 1).

The above described algorithm should give good results when $\int_0^1P_r(\xi){\rm d}\xi$ exists (cf.
Eq.~\ref{L}). In our case $P_r(\xi)$ can be approximated by the power function $(1-\xi)^{-2\beta}$ when
$\xi\rightarrow 1$. The exponent $\beta$, which comes from the approximation for $q_L(\xi)$, turns out to be
larger than one half (see discussion in the next section). Thus if the data were extrapolated to the contact
point $\xi=1$, this integral would diverge. It is well known, however, that the evaporation heat flux is
limited \cite{Carey} by a flux $q_{max}$. As calculated in the kinetic theory of gases
\begin{equation}\label{qmax}
q_{max}=0.74\rho_VH\sqrt{R_gT_{sat}/(2\pi M)} \approx 10^4\mbox{ MW/m}^2.
\end{equation}
In our model the above divergence appears because of the assumption that the temperature remains constant
along the vapor-liquid interface. In reality, this assumption is violated in the very close vicinity of the
contact point where the heat flux $q_L$ is comparable to $q_{max}$. Thus we accept the following
approximation for the function $q_L(\xi)$, $\xi<1$. It is extrapolated using the power law
$q_L(\xi)\propto(1-\xi)^{-\beta}$ until it reaches the value of $q_{max}$ and remains constant while $\xi$
increases to unity. This extrapolation is used to calculate the integrals in (\ref{Vn}) and (\ref{Prn}).
There is no need to modify the constant-temperature boundary condition for the heat transfer calculations
because the calculated heat flux $q_L$ remains always less than $q_{max}$.

The calculations show that the function $q_L(\xi)$ (see Fig.~\ref{q_L}) can be described well by the above
power law where $\beta\sim 1$ grows slightly with time. We note that for the growing bubble the divergence is
stronger than for the $90^{\circ }$ wedge analyzed in \cite{EuLet}. The difference between these two cases is
the behavior of the heat flux $q_{S}$ in the vicinity of the contact point. While it was supposed to be
uniform for the $90^{\circ }$ wedge, the function $q_{S}(x)$ increases strongly near the contact point for
the growing bubble case, see the discussion associated with Fig.~\ref{qT_S}.

Sometimes, an occasional ``bump" on the $q_L(\xi)$ curve appears during the iteration of the steps 3--7 of
the algorithm because of inaccurate calculation of the bubble shape when the automatically chosen number of
the boundary elements in the vicinity of the contact line is too small. This bump disappears during at most
three time steps. This disappearance indicates a good numerical stability of the algorithm. When the bubble
evolution is exceedingly slow, it may be necessary to increase the time step several times. While not
influencing accuracy strongly (this is an intrinsic property of the BEM \cite{Lagier}), such a change
decreases the temporal resolution.

\section{Results and discussion}\label{sec7}

\begin{figure} \centering\includegraphics[width=8cm]{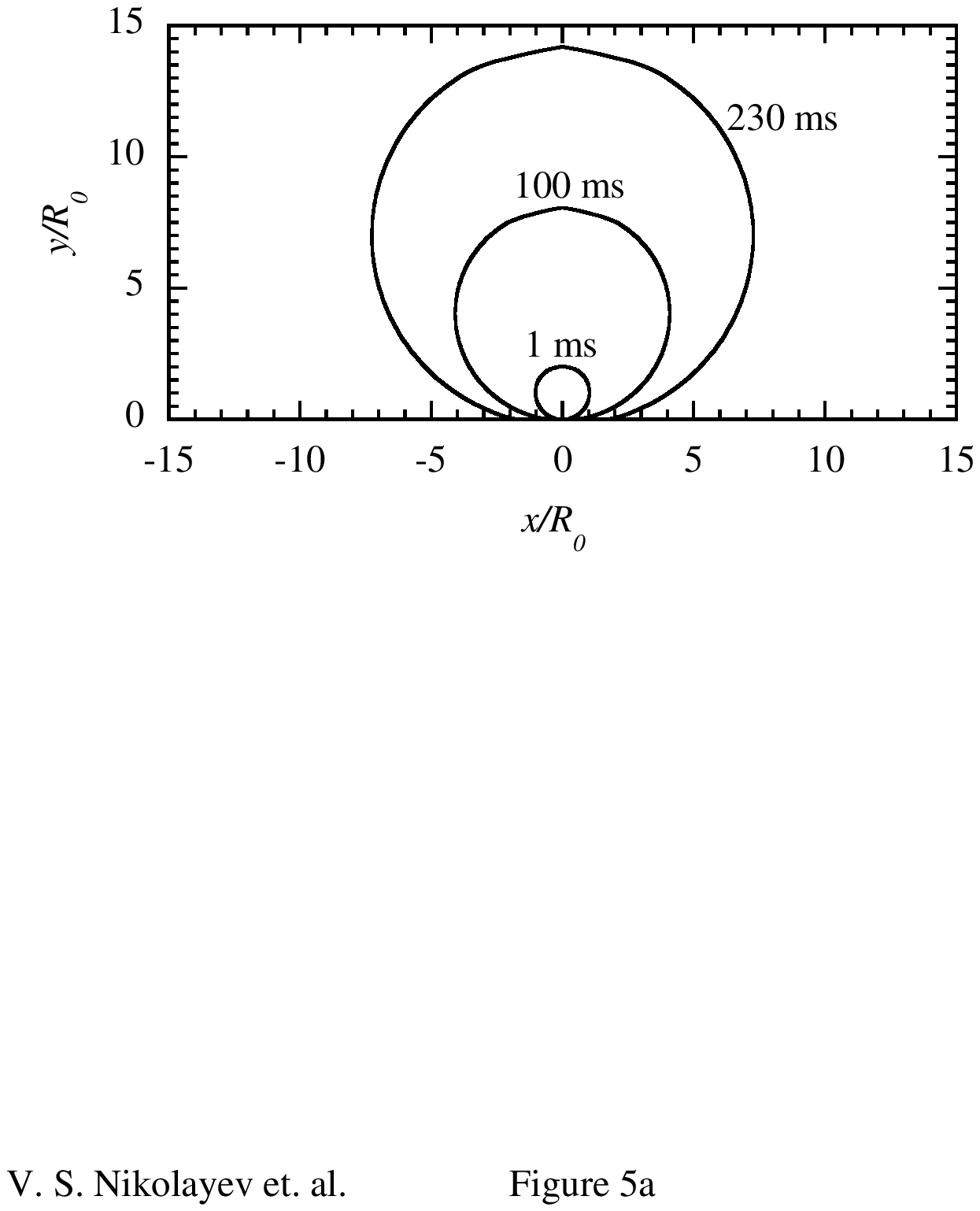}\includegraphics[width=8cm]{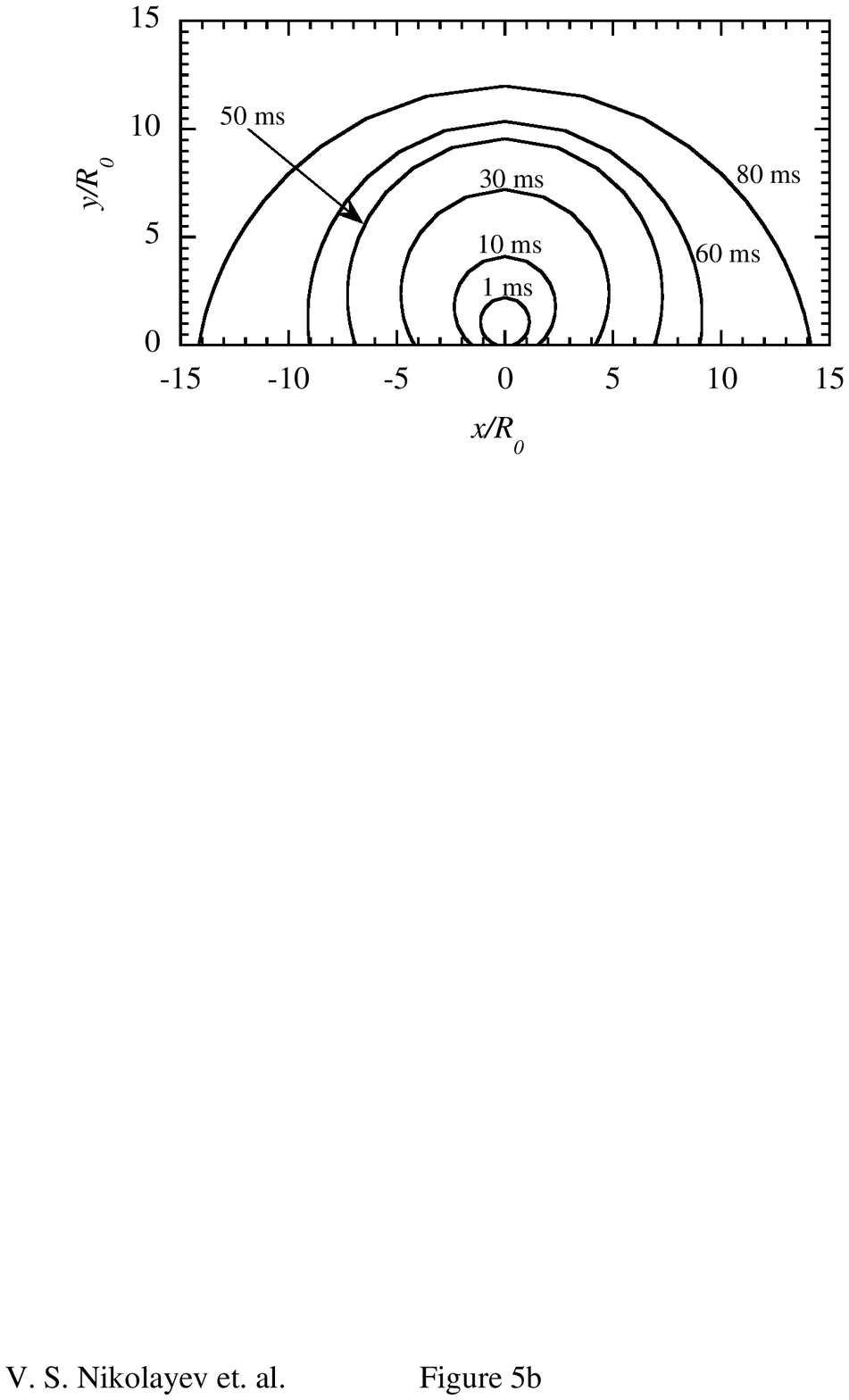}\caption{The bubble shape shown for the different growth times.
a)$q_0=0.05$ MW/m${}^2$; b) $q_0=0.5$ MW/m${}^2$;}\label{shape}
\end{figure}

\begin{figure}[ht]\centering\includegraphics[width=8cm]{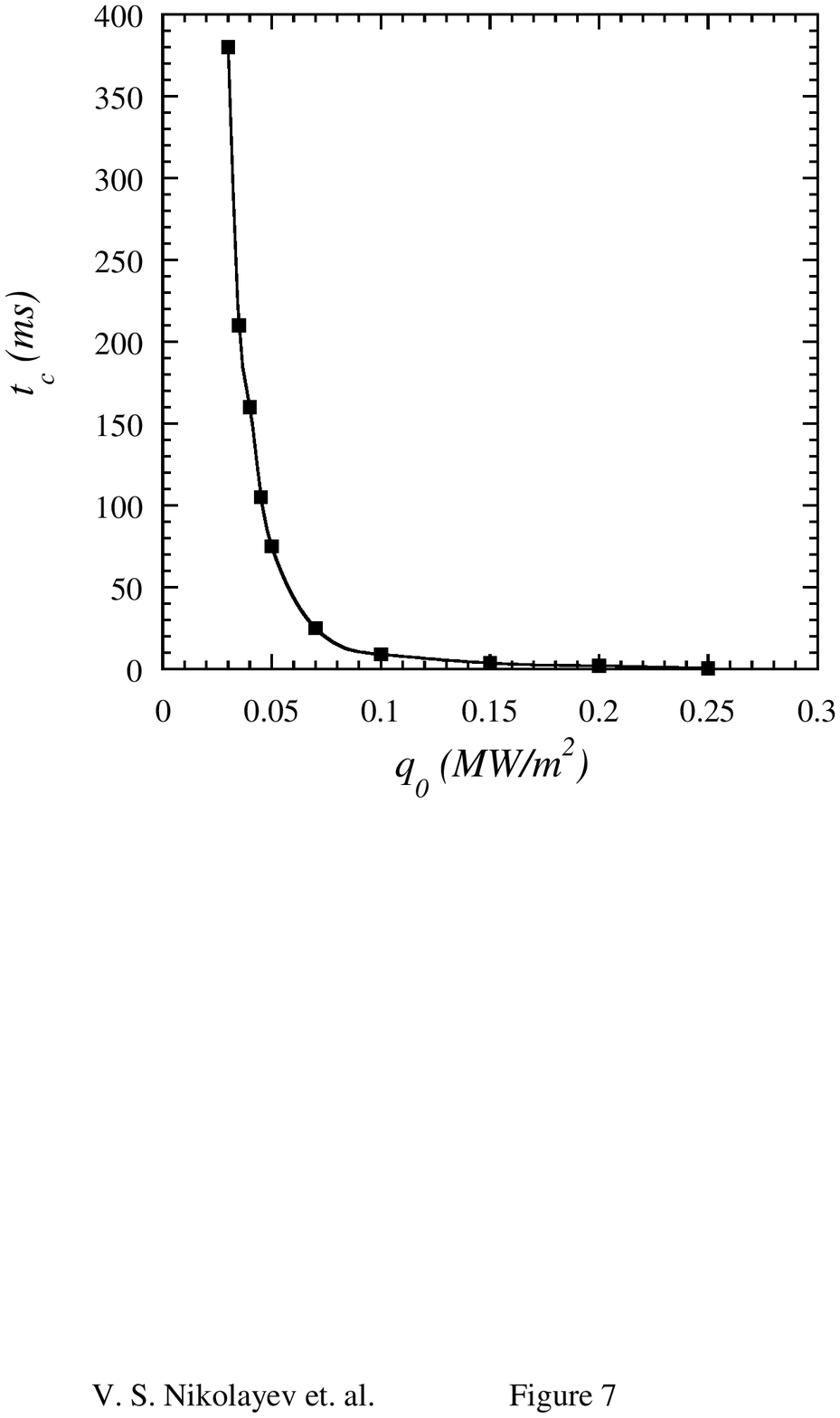}
\caption{The transition time $t_c$ as a function of the heat flux $q_0$.} \label{time}
\end{figure}

There is a number of new results that we have obtained from this simulation. The most important is the time
evolution of the dry spot under the vapor bubble. At low heat flux, the shape of the bubble stays nearly
spherical (Fig.~\ref{shape}a) until it leaves the heating surface under the action of gravity or hydrodynamic
drag forces. Fig.~\ref{shape}b shows that at large heat flux the radius of the dry spot can approach the
bubble radius during its evolution, thus confirming previous theoretical predictions \cite{EuLet} where the
apparent contact angle grow with time. It should be emphasized that the {\em actual} contact angle remains
zero during the evolution, see sec.~\ref{S3}. The large {\em apparent} contact angle is due to the strong
change in slope of the bubble contour near the contact line where the vapor recoil force is very large (see
\cite{EuLet} for the advanced discussion). The temporal evolution of the radius $R_d$ of the dry spot is
illustrated in Fig.~\ref{dry_t}, where the time evolution of $R_d/R$ is shown. $R$ is the visible bubble
radius defined as the maximum abscissa for the points of the bubble contour as shown in Fig.~\ref{bubble}.
Note that $R_d/R\leq 1$ by definition. At low heat fluxes $q_{0}<100$ kW/m${}^{2}$ $R_d$ stays very small
during a long time interval (see Fig.~\ref{shape}a). In this regime the bubble should leave the heater
quickly because of the small adhesion that is proportional to the contact line length. After a transition
time $t_{c}$ which depends on $q_{0}$, the growth of the dry spot accelerates steeply (see Fig.~\ref{dry_t}).
This time $t_{c}$ corresponds to the moment where the growing vapor recoil force becomes comparable to the
surface tension. This force balance was analyzed in details in \cite{EuLet}, where numerical estimates were
given. The dependence $t_{c}(q_{0})$ is presented in Fig.~\ref{time}. Clearly, $t_{c}$ is a decreasing
function of $q_{0}$. This means that, at a sufficiently large $q_0$, the dry spot becomes very large in a
short time and the departure of the bubble is prevented because of the large adhesion to the heater. During
the further growth this bubble can either create alone a nucleus for the film boiling or coalesce with
another similarly spreading neighboring bubble. Therefore, we can associate this value of $q_0$ with the
$q_{CHF}$. Without a careful analysis of the time of departure, it is not possible to determine a precise
value for $q_{CHF}$. This will be the subject of future studies.

We neglected the initial superheating for the sake of simplicity. The initial superheating would accelerate
the bubble growth slightly in the initial stages that are not important for the dry spot spreading that
becomes significant later on.

Slight oscillations in the dry spot growth are clearly visible in Fig.~\ref{dry_t}, especially in the fast
growth regime. We varied the numerical discretization parameters in order to check whether these oscillations
appear due to a numerical instability. Neither the amplitude nor frequency of the oscillations depends on
numerical discretization parameters. We conclude that the oscillations reflect a physical effect (see below)
rather than a numerical artifact.

The kinetics of the bubble growth is illustrated in Fig.~\ref{rad}, where the temporal evolution of the
bubble radius $R$ is presented. At $t<t_c$ we recover a general tendency of the bubble growth curves (see, e.
g. \cite{Cooper,Stephan}), where $R\propto t^{1/2}$. At $t>t_c$ the growth exponent is larger. The curve
$R(t)$ exhibits oscillations with their amplitude increasing with time. We suspect that this effect appears
when the temperature distribution in the heater responds too slowly to maintain the fast growth rate in the
bubble.

\begin{figure}[ht]\centering\includegraphics[width=8cm]{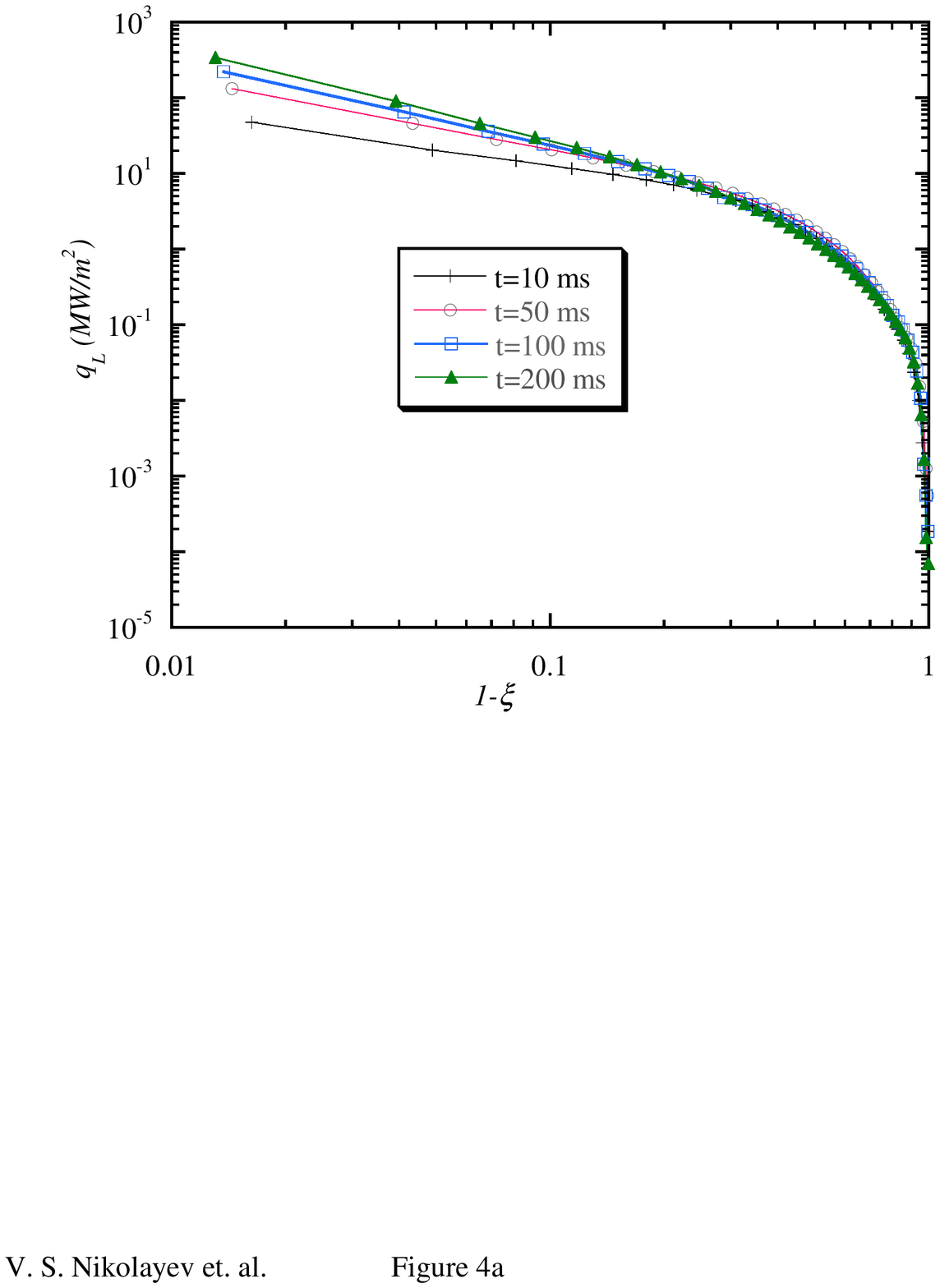}\includegraphics[width=8cm]{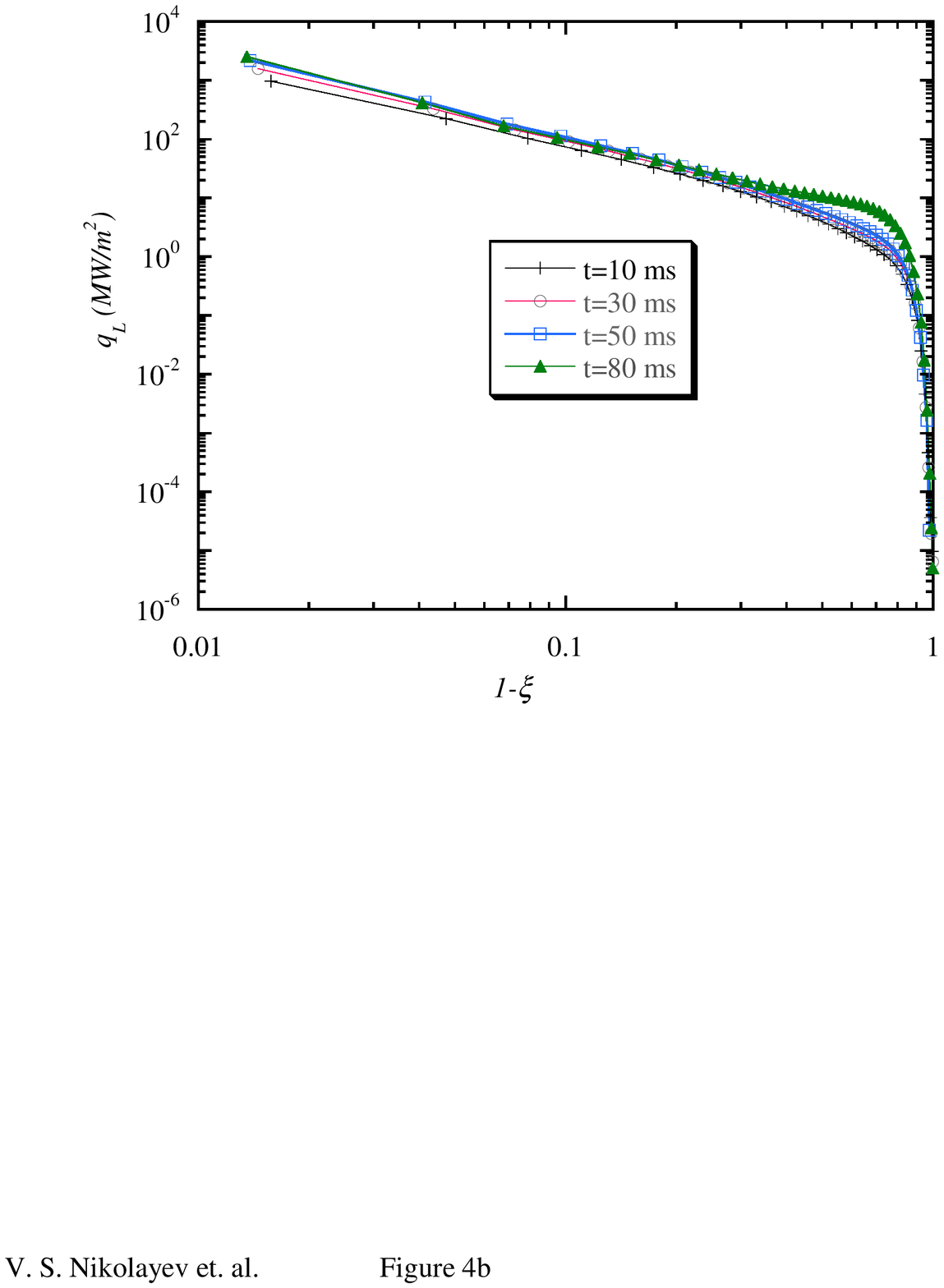}
\caption{Variation of the heat flux $q_L$ (defined in (\ref{qL})) along the bubble contour for different moments
of time. The curvilinear coordinate $\xi$ varies along the bubble contour; $\xi=1$ at the contact point: a)
$q_0=0.05$ MW/m${}^2$, b) $q_0=0.5$ MW/m${}^2$.} \label{q_L}
\end{figure}

\begin{figure} \centering\includegraphics[width=8cm]{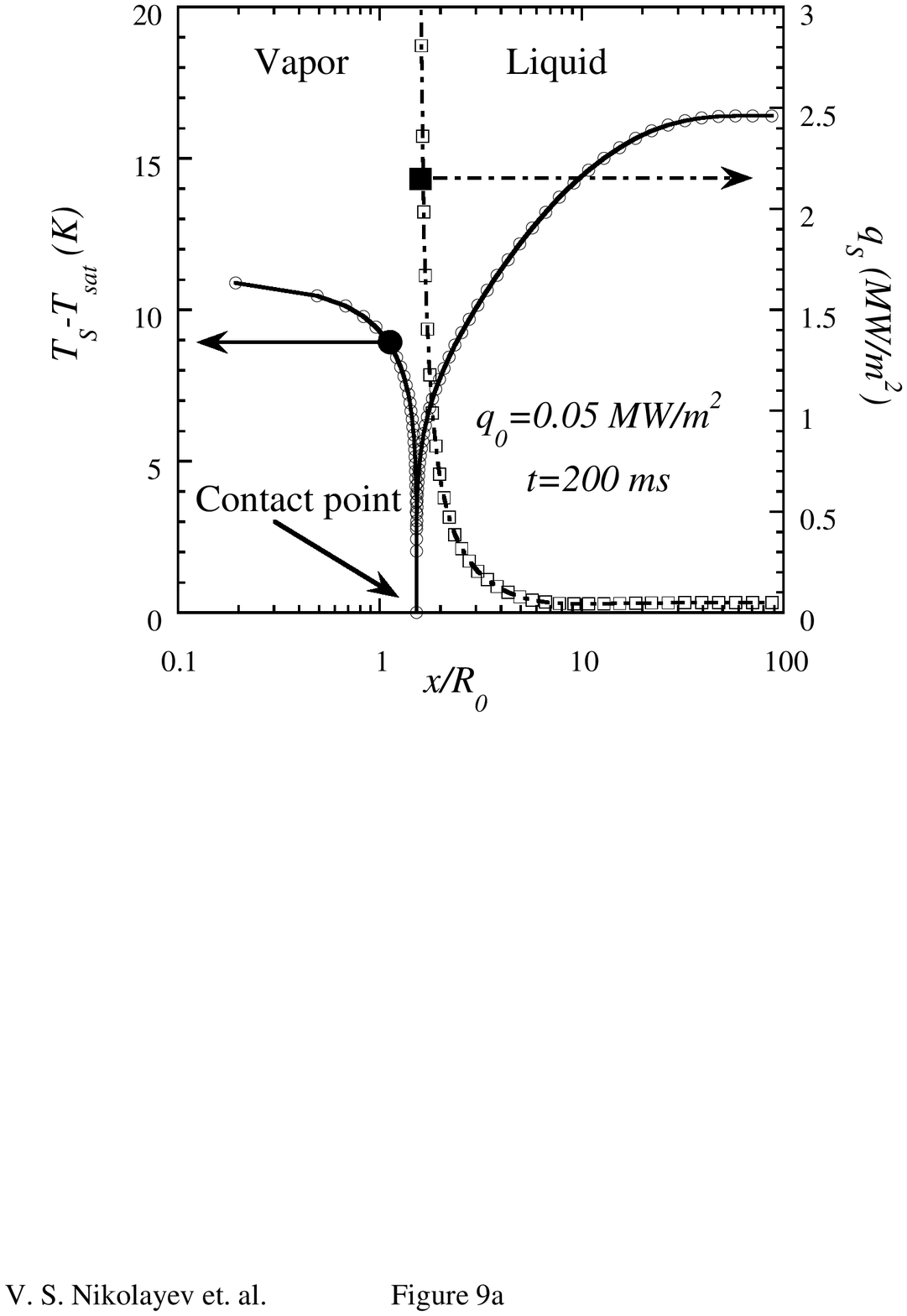}\includegraphics[width=8cm]{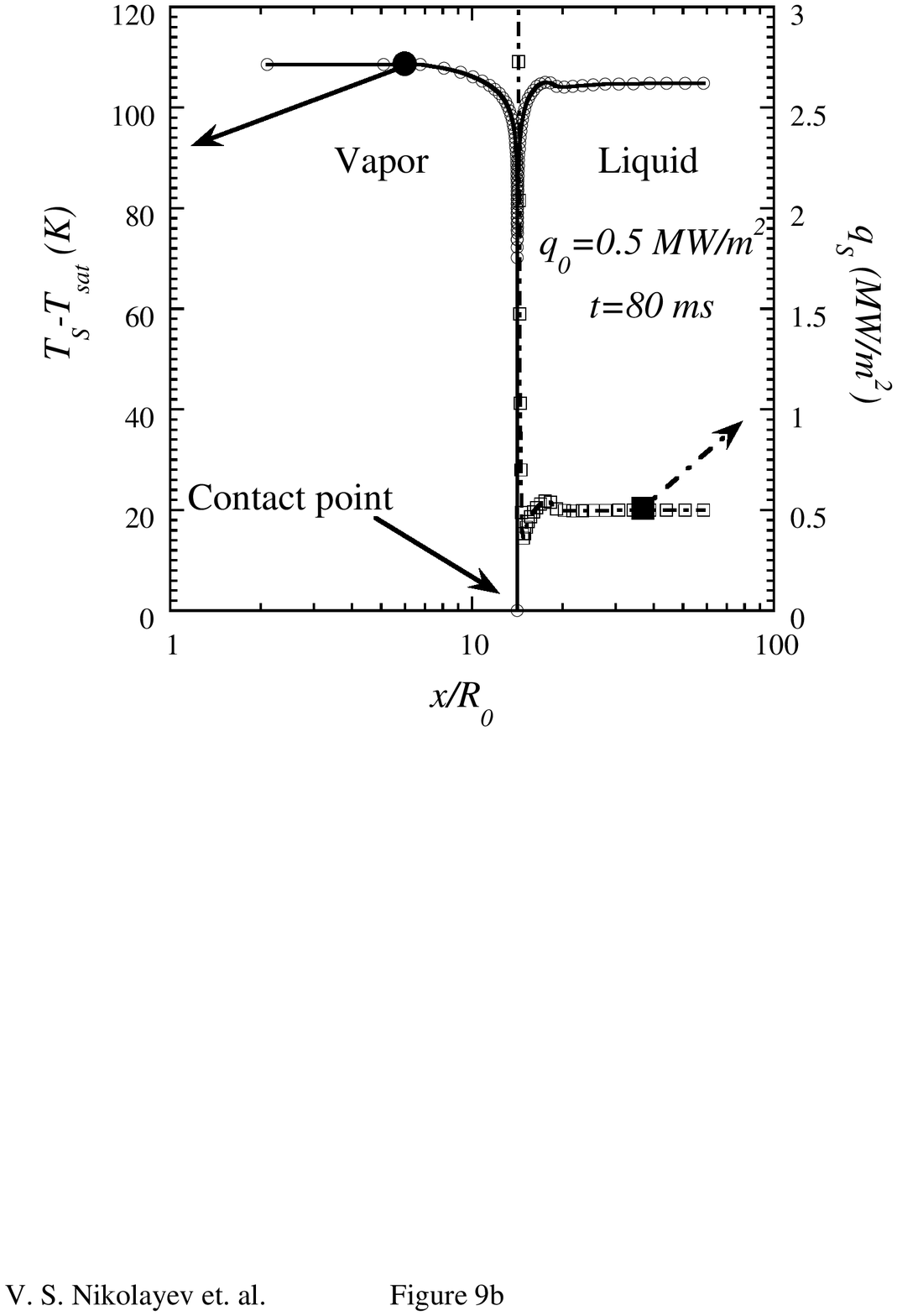}\caption{Variation of the heat flux $q_S$ and the temperature $T_S$ along the surface of the heater for  a)$q_0=0.05$ MW/m${}^2$, $t=200$ ms;
b)$q_0=0.5$ MW/m${}^2$, $t=80$ ms. The point $x=0$  corresponds to the center of the bubble. $T_S-T_{sat}=0$ at
the contact point, $q_S=0$ to the left of it, i.e. inside the dry spot.} \label{qT_S}
\end{figure}

Our simulation enables the heat transfer under the bubble to be rigorously calculated. The variation of the
heat flux $q_{S}$ along the heating surface is shown in Figs.~\ref{qT_S}a,b for the different values of the
heat flux $q_0$. The value of $q_{S}$ on the liquid side in the vicinity of the contact line turns out to be
very close to $q_{L}$, the heat flux that produces evaporation on the vapor-liquid interface and that
diverges on the contact line (see Figs.~\ref{q_L}). This correspondence was expected, because all the heat
flux supplied by the heater to the foot of the bubble is consummated to evaporate the liquid, in agreement
with the ``liquid microlayer" models. Since $q_{S}$ is zero (cf. (\ref{qS0})) if the contact line is
approached from the dry spot side, the function $q_{S}(x)$ is discontinuous in the vicinity of the contact
point. Far from the bubble $q_{S}=q_{0}$ as it should be.

The variation of the temperature along the heating surface $T_S(x)$ is also shown in Figs.~\ref{qT_S}. Far
from the bubble, $T_S$ has to increase with time independently of $x$ and follows a square root law according
to (\ref{TSinf0}). It decreases to $T_{sat}$ near the contact point because the temperature should be equal
to $T_{sat}$ on the whole vapor-liquid interface, according to the imposed boundary condition.
Figs.~\ref{qT_S} demonstrate that there is a zone of lowered temperature around the bubble, in agreement with
the experimental observations \cite{Cooper}. Inside the dry spot, $T_{S}$ increases with time sharply because
the heat transfer through the dry spot is blocked. Fig.~\ref{qT_S}b shows that at high heat flux and
$t>t_{c}$ the temperature inside the dry spot becomes larger than the temperature far from the bubble. This
temperature increase leads to eventual burnout of the heater observed during the boiling crisis. The presence
of singularities in the functions $T_S(x)$ and $q_{S}(x)$ is the reason for our choice of the boundary
conditions on the heating surface in the form (\ref{qS0},\ref{TS0}). As a matter of fact, an application of
the conditions of the uniform heat flux or uniform temperature would lead to the physically inconsistent
results such as non-integrable divergency of $q_{L}$ at the contact line. We note that an error in the
calculation of $q_{L}$ should strongly influence the results for the dry spot dynamics.

\section{Conclusions}\label{sec8}

The 2D free-boundary simulation allows us to calculate the actual bubble shape and the variation of the
temperatures and fluxes along the vapor-liquid, vapor-solid, and liquid-solid interfaces. The description of
the heat transfer in the vicinity of the triple contact line presents the most difficult part of the problem.
Our variation of the Boundary Element Method is capable of adequately describing it.

Our main result is the evidence for the growth of the dry spot under the vapor bubble. While this increase is
very slow in the beginning of the bubble growth, it accelerates steeply after a growth time $t_c$ that
depends on the external heat supply. At low heat supply, $t_c$ is very large so that the bubble can grow
large enough to satisfy the conditions for departure from the heating surface {\em before} the dry spot
becomes significant. In contrast, at high heat supply, $t_c$ is small so that the dry spot grows very rapidly
which means that the bubble spreads over the heating surface. Although our analysis is limited to the case of
high system pressures, we note that at low pressures this effect can also be important because the forces of
dynamical origin ``press'' the bubble against the heater, thus favoring its spreading. The results of this
simulations thus confirm the validity of the ``drying transition'' model suggested in \cite{EuLet} to
describe the boiling crisis.

Unfortunately, observations of the  bubble shape and the dry spot growth during boiling at high pressures and
high heat fluxes are unknown to us, preventing a direct comparison of our results with the experimental data.
We note, however, that the growth of the dry spot immediately before the boiling crisis was observed in
\cite{Van,Tor}, where observations have been carried out through a transparent heating surface. The authors
of \cite{Tor} state that ``When the heat flux is sufficiently large, suddenly at some point on the heating
surface a dry area is not wetted and starts growing, leading to burnout". This observation confirms directly
the validity of our model.

\subsection*{Acknowledgments}

The authors would like to acknowledge the financial support of EDF (V.~N. and D.~B.) and NASA (V.~N. and
J.~H.). This work has been done when one of the authors (V.~N.) stayed at the Department of Physics of the
UNO, which he would like to thank for its kind hospitality. V.~N. thanks Dr. Herv\'e Lemonnier for the
introduction into the BEM for the heat diffusion. The authors are grateful to Jean-Marc Delhaye for fruitful
discussions.

\newpage
\begin{table}[ht]
\begin{tabular}{|c|c|c|c|}\hline
 Description & Notation & Value & Units \\ \hline
  Saturation temperature & $T_{sat}$ & 311 & ${}^\circ$C \\
  Thermal conductivity of liquid & $k_L$  & 0.55 & W/(m K) \\
  Specific heat of liquid& $c_{pL}$ & 6.12 & J/(g K) \\
  Mass density of liquid& $\rho_L$ & 688.63 & kg/m${}^3$ \\
  Mass density of vapor& $\rho_V$ & 55.48 & kg/m${}^3$ \\
  Latent heat of vaporization & $H$ & 1.3 & MJ/kg \\
  Surface tension &$\sigma$  & 12.04 & mN/m \\
  Thermal conductivity of steel & $k_S$  & 15 & W/(m K) \\
  Specific heat of steel& --- & 0.5 & J/(g K) \\
  Mass density of steel& --- & 8000 & kg/m${}^3$ \\
  Initial bubble radius&$R_0$&0.05&mm\\
  Reference heat flux&$\bar{q}$&1&MW/m${}^2$\\
  Reference thermal conductivity & $\bar{k}$  & 1 & W/(m K) \\
 Minimal discretization step &$d_{min}$&0.001&$R_0$\\  Time step&$\Delta t$&1&
 ms\\ \hline
\end{tabular}
\caption{Values of parameters used in the simulation.}
\end{table}

\begin{figure}[ht]\centering\includegraphics[width=8cm]{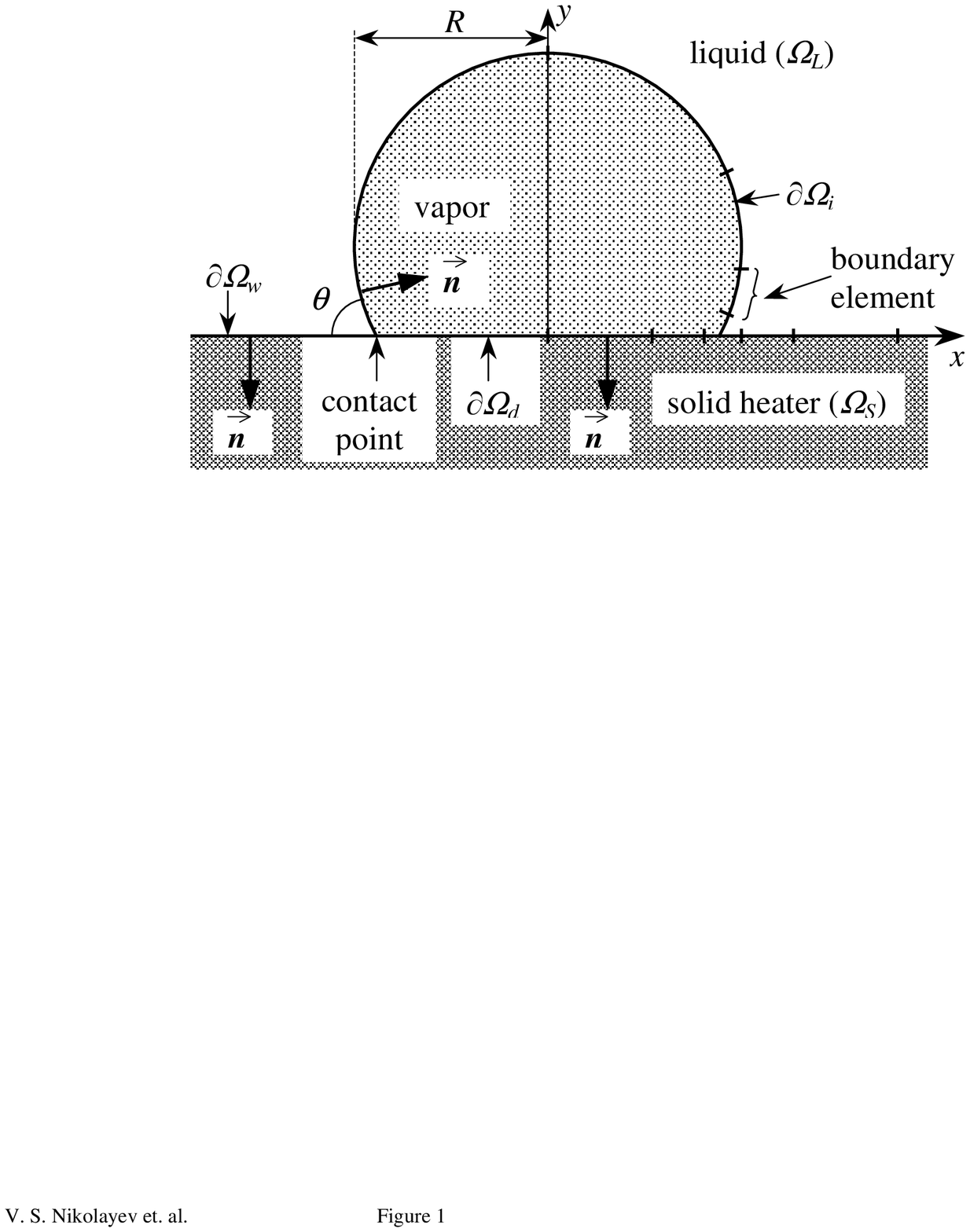}
\caption{Vapor bubble on the heating surface surrounded by liquid. The chosen direction of the unit normal
vector $\vec{n}$ is shown for each of the subcontours $\partial\Omega_w$, $\partial\Omega_d$ and
$\partial\Omega_i$. The discretization is illustrated for the right half of the subcontours.} \label{bubble}
\end{figure}

\begin{figure}[ht]\centering\includegraphics[width=8cm]{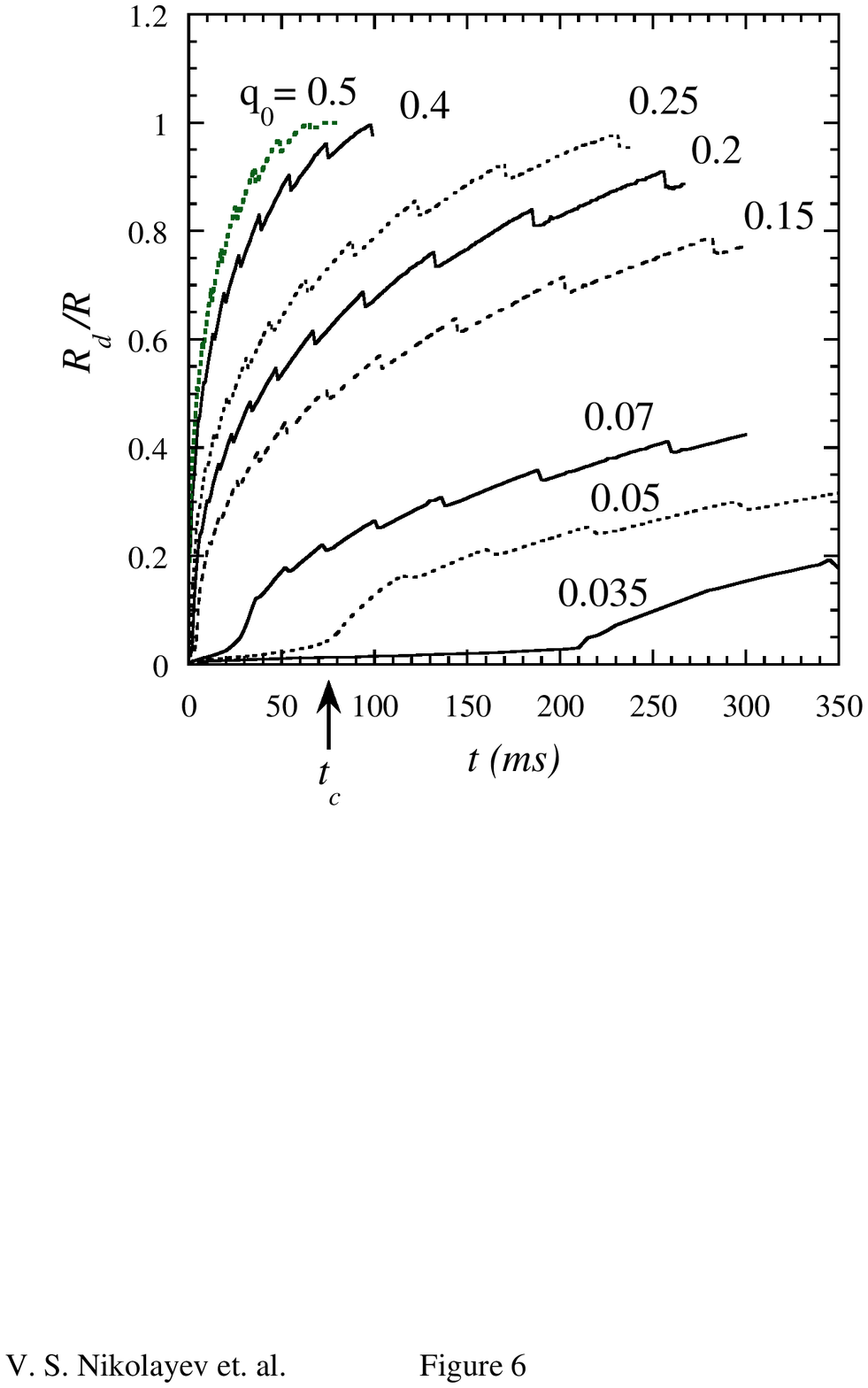}
\caption{The temporal evolution of the quotient of the dry spot radius $R_d$ and the bubble radius $R$ for
different values of $q_0$ expressed in MW/m${}^2$. $R$ is measured as shown in Fig.~\ref{bubble}. The transition
time $t_c$ is shown for $q_0=0.05$ MW/m${}^2$.} \label{dry_t}
\end{figure}
\begin{figure}[ht]\centering\includegraphics[width=8cm]{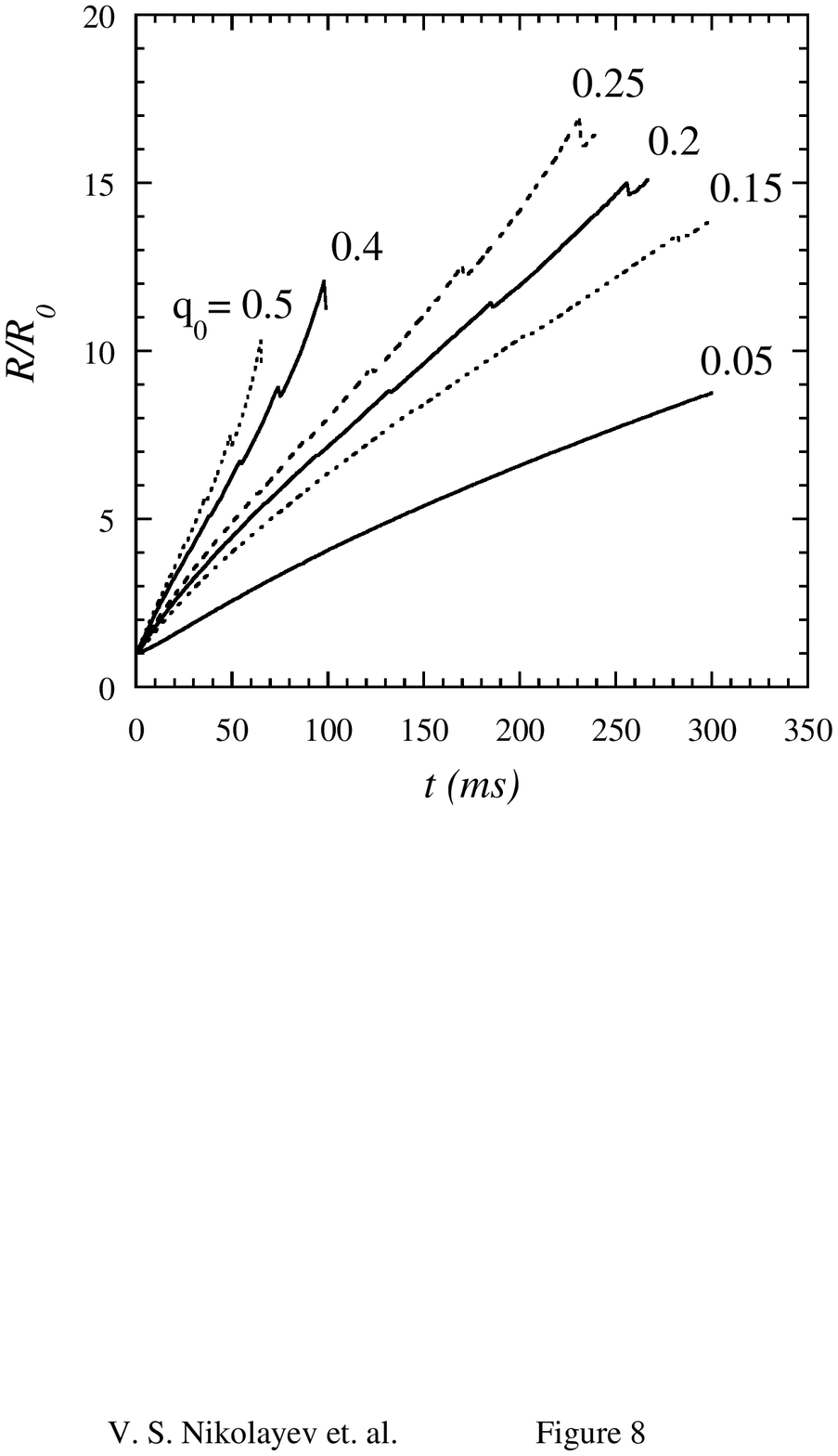}
\caption{Temporal evolution of the bubble radius $R$ for different values of $q_0$ expressed in MW/m${}^2$. $R$
is measured as shown in Fig.~\ref{bubble}.} \label{rad}
\end{figure}
\appendix
\section{Volume determination}\label{A1}

The volume $V$ of an object $\Omega$ can be calculated as
\begin{equation}
V=\int\limits_{(\Omega)} {\rm d}\Omega={1\over 2}\int\limits_{(\Omega)}
\mbox{div}(x\vec{e}_x+y\vec{e}_y)\;{\rm d}\Omega,\label{V1}
\end{equation}
using the obvious equality $$\mbox{div}(x\vec{e}_x+y\vec{e}_y)=2,$$ where $\vec{e}_x=(1,0)$ and
$\vec{e}_y=(0,1)$ are the unit vectors directed along the axes. The Gauss integral theorem is valid for any
$\vec{a}$ and $\Omega$:
\begin{equation}
\int\limits_{(\Omega)}\mbox{div}\vec{a}\;{\rm d}\Omega =\int\limits_{(\partial\Omega)} \vec{a}\cdot\vec{n}_e
\;{\rm d}\,\partial\Omega, \label{Gauss}
\end{equation}
where $\partial\Omega$ denotes the surface of $\Omega$, and $\vec{n}_e$ is the external unit normal vector to
this surface. In our case $\partial\Omega=\partial\Omega_i\cup\partial\Omega_d$, where $\partial\Omega_d$ is
the surface of the vapor-solid contact, i.e. the dry spot. The application of the equality (\ref{Gauss}) to
the last integral in (\ref{V1}) yields the expression
\begin{equation}
V={1\over 2}\int\limits_{(\partial\Omega_i\cup\partial\Omega_d)} (xn_e^x+yn_e^y) \;{\rm
d}\,\partial\Omega.\label{V2}
\end{equation}
Since $y=n_e^x=0$ on $\partial\Omega_d$ (see Fig.~\ref{bubble}), the integral over it is equal to zero. Thus
(\ref{V2}) reduces to (\ref{V}).

\end{document}